# THERMAL MODELING OF HIGH POWER LED MODULES

*D.A. Benoy*

Philips Lighting
P.O. Box 80200, 5600 JM Eindhoven, The Netherlands

## ABSTRACT

This paper presents a study of accuracy issues in thermal modeling of high power LED modules on system level. Both physical as well as numerical accuracy issues are addressed. Incorrect physical assumptions may result in seemingly correct, but erroneous results. It is therefore important to motivate the underlying key physical assumptions of a thermal model. In this paper thermal measurements are used to calibrate a computational fluid dynamics (CFD) model of a high power LED module model at a reference application condition, and to validate it at other application conditions.

## 1. INTRODUCTION

One of the key advantages of LED's are the high energy and optical system efficiencies and the product design freedom, due to their small form factor. Some illumination applications require white high power LED modules under a broad and versatile range of ambient boundary conditions. A prototype of a passively cooled high power density LED module is shown in picture A and B of Figure 1. For these applications thermal management is a major issue for both optical and reliability properties of the LED module shown in Figure 2.

For thorough analysis of the thermal performance, and further optimization of these high LED modules a detailed thermal model has been developed for performing thermal simulations. For future design and product development work it is also important to know how to perform predictive thermal model calculations when prototype modules are not yet available. This implies that the accuracy of a thermal model much be such that it becomes predictive, or that the source of model inaccuracy can be identified. This paper presents a study of accuracy issues in thermal modeling of these high power LED modules. Thermal measurements are used to calibrate and validate a thermal computational fluid dynamics (CFD) model by comparing temperatures profiles of various components of the module. The IcePak™ software package is used to implement a CFD thermal model on system level, which is shown in Figure 2. The CFD thermal comprises the whole LED module including the geometric details of the internal material interface layers, the heat sink geometry and environmental boundary conditions.

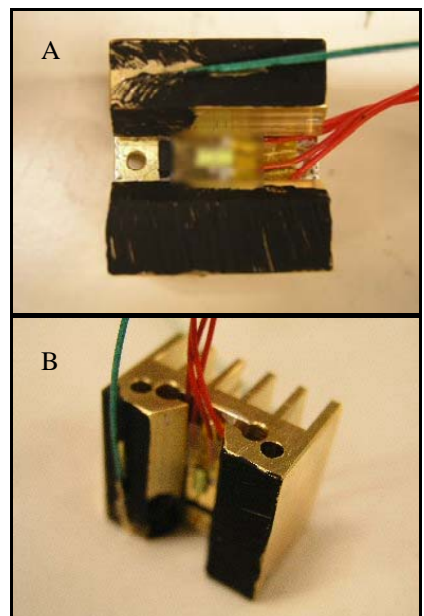

**Figure 1.** The pictures show the top and isometric view of a high power LED module. Some parts are painted black for IR-measurements.

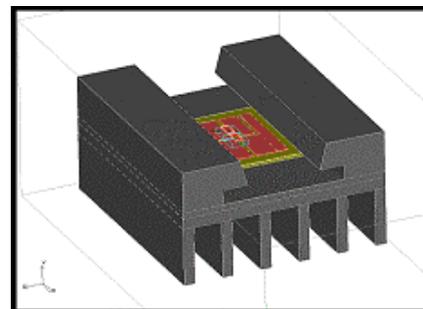

**Figure 2.** IcePak™ solid model of the high power LED module.

## 2. THERMAL MODELING

The finite volume IcePak™ software package is used to perform package level thermal model calculations of the LED module shown in Figure 2. It is well known that both physical and numerical issues have a large impact on the accuracy of the computed results [1]. Both issues are considered in the calibration and validation phases of the CFD thermal model of the high power LED module. Concerning physical issues CFD packages requires 1) the specification of the flow regime: i.e. laminar, or turbulent flow, 2) thermal radiation properties, and 3) specification of heat generation in heat dissipating components. Concerning the specification of the heat generation a part of the input electric power is converted into light, which is described by the wall plug efficiency (*WPE*) of the LED components. The *WPE* is defined as the ratio of the optical visible and electrical power and is an important characteristic performance number for LED devices. The heat dissipation in the LED's is then given by $P_{heat} = I \times V \times (1-WPE)$, where the *WPE* must be treated as an input parameter for the model calculations. The *WPE* is thus an additional source of uncertainty when it comes to a quantitative comparison between thermal CFD model calculations and experimental data.

Different types of model calculations are performed:
- Laminar and turbulent flow,

In case of a turbulent flow regime turbulence is described by the zero-equation approach. Wall plug efficiencies of 10% and 15% are assumed and the electric input powers of the LED modules are 5.25 and 8.26W.

A combination of errors in the aforementioned physical issues may result in temperature distributions and temperature values at different module components, which are seemingly in good agreement with thermal measurements. For example the assumptions of a turbulent flow regime including thermal radiation with a "too low" *WPE* of 10% and an electric input power of 8.26W result in a temperature distribution and component temperatures at the die, the ceramic substrate, and the heat sink, which agree well with measured values. In case of IcePak™'s default grid settings the maximum temperature differences between measured and calculated temperatures is 4°C at the die (measured 169°C) and the heat sink (measured 124°C, see Table 1). Average surface temperatures of the high power LED module are measured with an IR-camera. The emissivities of the various different materials are calibrated at an elevated temperature of 40°C.

| Component | Calculated temperature | Measured temperature |
|---|---|---|
| LED chip | 173.0°C | 169.0°C |
| Submount | 131.6°C | 134.1°C |
| Ceramic substrate | 127.6°C | 129.0°C |
| Heat sink | 127.2°C | 124.0°C |

**Table 1. Calculated versus measured temperatures at different module components. The flow is assumed to be turbulent, the *WPE* = 10%, and thermal radiation is included. For an electric power input of 8.26W this implies a power dissipation of 7.46W.**

But is the assumption of a turbulent flow justified? The nature of the natural convection flow is predicted by the dimensionless Rayleigh number $Ra = Gr \times Pr$, where *Gr* and *Pr* are the Grashof and Prandtl numbers, respectively. The Grashof number is defined as

$$Gr = \frac{\beta g \Delta T L^3}{\nu^2}$$

where $\beta$ is the thermal expansion coefficient of air, *g* is the gravitational constant. $\Delta T$ is temperature difference between LED module and ambient temperature. For a safe estimation the largest possible $\Delta T$ is considered by taking a typical die temperature for the LED module temperature. *L* and $\nu$ are a characteristic module length and the kinematic viscosity coefficient, respectively. The Prandtl number is defined as

$$Pr = \frac{\nu}{\kappa}$$

where $\kappa$ is the thermal diffusion coefficient. For the current application the approximate values for the Grashof and Prandtl numbers are

$$Gr = \frac{0.0033[\text{K}^{-1}] \times 9.81[\text{m/s}^2] \times 100[\text{K}] \times 0.03^3[\text{m}^3]}{(1.60 \times 10^{-5})^2 [\text{m}^4/\text{s}^2]}$$

$$= 3.4 \times 10^5$$

$$Pr = \frac{1.60 \times 10^{-5}[\text{m}^2/\text{s}]}{2.3 \times 10^{-5}[\text{m}^2/\text{s}]} = 0.70$$

so that

$Ra = 3.4 \times 10^5 \times 0.70 = 2.38 \times 10^5$.

According to [2] transition to turbulence in buoyant flows in vertical enclosures occurs at $Ra \sim 4 \times 10^6$ (for enclosures with aspect ratios of the order 8). This critical value is much larger than the estimated Ra value ($2.38 \times 10^5$) for the LED module so that the flow regime is expected to be laminar. This means that if we do include turbulence in the thermal calculations the heat transfer will be overestimated and the temperatures of the LED module will be too low.

Switching from turbulent to laminar flow, and do the IcePak™ calculation again then the maximum differences between measured and calculated temperatures increases up to 12°C at the die and the heat sink (see Table 2). These are systematic temperature differences of the order of 10%.

| Component | Calculated | Measured |
|---|---|---|

|  | temperature | temperature |
|---|---|---|
| LED chip | 181.0°C | 169.0°C |
| Submount | 140.8°C | 134.1°C |
| Ceramic substrate | 136.6°C | 129.0°C |
| Heat sink | 136.0°C | 124.0°C |

**Table 2. Calculated versus measured temperatures for a laminar flow, including thermal radiation, and *WPE* = 10%.**

The observation that the calculated temperatures at all module components are systematically too high brings us to the plausible conclusion that a *WPE* of 10% is too low. If the LED components are more efficient than anticipated, the *WPE* must then exceed 10%. Increasing the *WPE* from 10 to 15% will result in a decreased heat dissipation from 90%×8.26W = 7.46W to 85%×8.26W = 7.00W. Thermal model calculations with a laminar flow, thermal radiation, and *WPE* = 15% result in maximum temperature differences between measured and calculated temperatures of 4°C, and 7°C at the die and heat sink, respectively.

| Component | Calculated temperature | Measured temperature |
|---|---|---|
| LED chip | 173.0°C | 169.0°C |
| Submount | 134.4°C | 134.1°C |
| Ceramic substrate | 131.6°C | 129.0°C |
| Heat sink | 131.0°C | 124.0°C |

**Table 3. Calculated versus measured temperatures for a laminar flow, including thermal radiation, and *WPE* = 15%.**

With the satisfactory agreement between calculated and measured temperature values we have in fact calibrated the thermal model calculation by fitting the *WPE* to a value of 15% resulting in a heat dissipation of 7.00W. By comparing the temperature data measured at electric input power of 5.25W with calculated results using the same *WPE* (15%) resulting in a heat dissipation power of 85%×5.25W=4.46W the thermal CFD model is validated. This comparison yields maximum temperature differences of 6.7% at the hottest module parts (i.e. the die) and 6.4% at the coolest module part (the heat sink) so that the thermal CFD model for this high power LED module is indeed validated.

| Component | Calculated temperature | Measured temperature |
|---|---|---|
| LED chip | 127.0°C | 118.3°C |
| Submount | 101.9°C | 101.3°C |
| Ceramic substrate | 99.6°C | 97.0°C |
| Heat sink | 99.0°C | 94.0°C |

**Table 4. Validation of calculated temperatures for a laminar flow, including thermal radiation, and *WPE* = 15%. The power dissipation in LED module is 4.46W.**

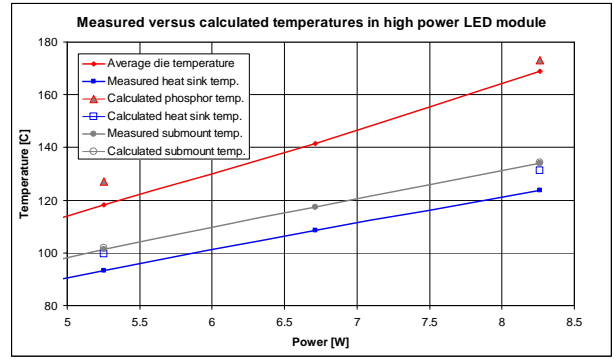

**Figure 3. Comparison between calculated and measured temperatures of the LED chips, substrate (top surface), and heat sink. The comparison is done for 2 powers (experimental loads of 5.25, and 8.26W).**

The calculated LED chip and heat sink temperature are both overestimated (7.6% and 5.6%, respectively) at the validation condition of *P*=5.25W. For the submount there is a very good agreement between the calculated and measured temperatures.

The thermal model calculations have been done with the default coarse mesh settings of IcePak$^{TM}$. The resulting grid near the edges of the fins of the heat sink is shown in Figure 4A. It is well known that in terms of solution accuracy a denser mesh is more accurate than a coarse mesh. When the laminar flow case including thermal radiation with a heat dissipation of 7.00W, i.e. a *WPE* of 15%, is solved using a finer meshes near the heat sink boundaries as shown in Figure 4B, then there is almost a perfect match between the measured and the calculated temperatures at the die (hottest module part) and ceramic substrate. Calculated temperatures of the main components of the LED module using a fine mesh are given in the third column of Table 5. For these LED module parts the differences are of the order of 0.5%, while for the heat sink the difference is 4%.

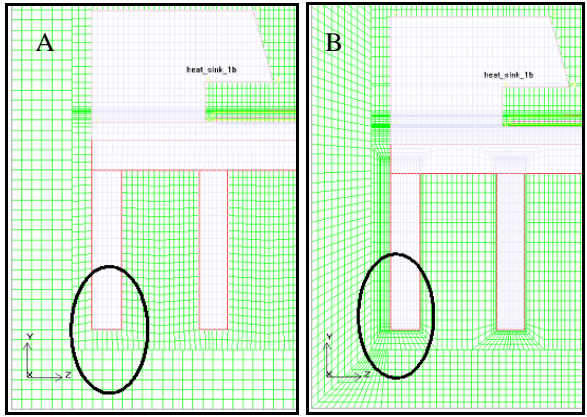

**Figure 4.** Default coarse (A), and fine (B) calculation mesh near the edges of the fins of the heat sink.

| Component | Coarse grid | Fine grid |
|---|---|---|
| LED chip | 173.0°C | 169.7°C |
| Submount | 134.4°C | 133.1°C |
| Ceramic substrate | 131.6°C | 129.4°C |
| Heat sink | 131.0°C | 128.5°C |

**Table 5.** Influence of mesh size on the calculated component temperatures.

### 2.1. Wall plug efficiency

In the thermal model calculations it was assumed that the LED components have a *WPE* of 15%. The question arises whether this *WPE* is consistent with available data of *WPE* of the used LED components. It is important to note that we can only estimate the *WPE* of these components. From luminous efficacy measurements the *WPE* is determined to be ±10%, which is considerably lower than the *WPE* = 15% used in the thermal model calculations. The discrepancy between the optically estimated *WPE*, and the *WPE* used in the thermal model illustrates that the current thermal model for CFD calculations has a limited accuracy.

### 3. CONCLUSIONS

This study gives us confidence in how to perform thermal model calculations of future LED modules. Still an unknown, but important parameter in the thermal modeling of LED modules is the wall plug efficiency (*WPE*) of the LED components, which does not follow from the thermal CFD calculations. From the thermal point of view the *WPE* of the LED components can be fitted when measured temperature results are available. However, from optical measurements the *WPE* can be estimated and is lower than the thermally estimated *WPE*.

For the analysis and optimization of the thermal performance of LED modules the current thermal model is good enough.